\documentclass{rnaastex}

\begin{document}

\title{Transient co-orbitals of Venus: an update}

\correspondingauthor{Carlos~de~la~Fuente~Marcos}
\email{nbplanet@ucm.es}

\author[0000-0003-3894-8609]{Carlos~de~la~Fuente~Marcos}
\affiliation{Universidad Complutense de Madrid \\
             Ciudad Universitaria, E-28040 Madrid, Spain}

\author[0000-0002-5319-5716]{Ra\'ul~de~la~Fuente~Marcos}
\affiliation{Universidad Complutense de Madrid \\
             Ciudad Universitaria, E-28040 Madrid, Spain}


\section{} 

Venus has no known satellites \citep{2009Icar..202...12S}, but has four known co-orbitals: (322756)~2001~CK$_{32}$ \citep{2004Icar..171..102B}, 
2002~VE$_{68}$ \citep{2004MNRAS.351L..63M,2012MNRAS.427..728D}, 2012~XE$_{133}$ \citep{2013MNRAS.432..886D}, and 2013~ND$_{15}$ 
\citep{2014MNRAS.439.2970D}. These objects are temporarily trapped in a 1:1 mean motion resonance with Venus, but are not gravitationally bound 
to it. It is believed that any putative primordial Venus co-orbitals (for example, Trojans) were lost early in the history of the Solar System;
present-day Venus co-orbitals are expected to be of transient nature \citep{2006Icar..185...29M}. Venus co-orbitals are very challenging targets 
as they spend most of the time in the unobservable (daytime) sky, at solar elongations well below 90\degr. The minimum orbit intersection 
distances (MOIDs) with the Earth of 322756, 2002 VE$_{68}$, 2012~XE$_{133}$, and 2013~ND$_{15}$ are 0.076, 0.027, 0.0019, and 0.0078 AU, 
respectively. This property alone makes them objects of significant practical interest as they tend to approach the Earth from the daytime side.

Here, we present numerical evidence suggesting that 2015~WZ$_{12}$ \citep{2015MPEC....X...04C} is a possible Venus co-orbital. With these 
tentative results we hope to encourage a search for precovery images of this minor body and perhaps even follow-up observations that may help in 
improving its poorly determined orbit so its current dynamical nature is better understood. The orbit determination of 2015~WZ$_{12}$ currently 
available (epoch JD 2458000.5) is based on 71 observations (1 Doppler) for a data-arc span of 6 d and has semi-major axis, $a$ = 0.721826$\pm$0.000012~AU, 
eccentricity, $e$ = 0.41250$\pm$0.00003, inclination, $i$ = 3\fdg6261$\pm$0\fdg0005, longitude of the ascending node, $\Omega$ = 251\fdg7613$\pm$0\fdg0006, 
and argument of perihelion, $\omega$ = 345\fdg7200$\pm$0\fdg0005.\footnote{\href{http://ssd.jpl.nasa.gov/sbdb.cgi}{JPL's Small-Body Database}} 
Unfortunately, this Aten asteroid can experience close encounters with Mercury, Venus and the Earth--Moon system, making its orbital evolution 
very chaotic; its MOID with the Earth is 0.0043 AU. With an absolute magnitude of 26.3 it may have a probable diameter of 19 m; large enough 
to cause a local disturbance, not too different from that of the Chelyabinsk event, in case of impact.  

\begin{figure}[h!]
\begin{center}
\includegraphics[scale=0.37,angle=0]{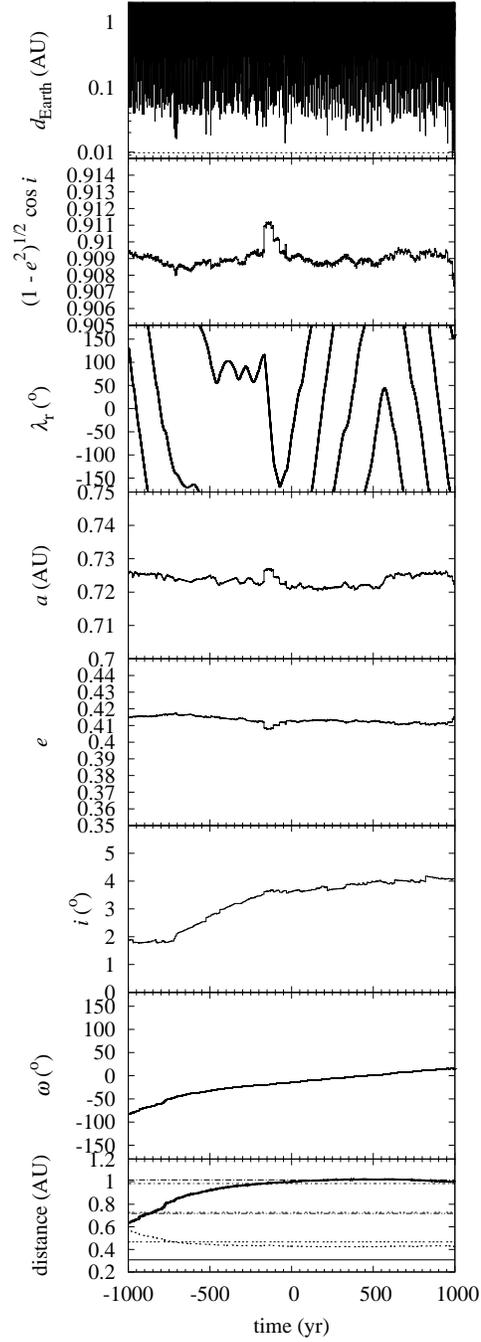}
\caption{Evolution of the values of the orbital elements and other relevant parameters for the nominal orbit of 2015~WZ$_{12}$.
         The top panel shows the geocentric distance (Hill radius of the Earth, 0.0098~AU). The Kozai-Lidov parameter is shown 
         in the second to top panel. The value of the resonant angle is displayed in the third to top panel. The evolution of 
         the orbital elements, semi-major axis, eccentricity, inclination and argument of perihelion is shown in the fourth to 
         top panel and the fourth, third and second to bottom panels, respectively. The bottom panel shows the distance from 
         the Sun to the descending (thick line) and ascending nodes (dotted line); the aphelion and perihelion distances of 
         Mercury, Venus and the Earth are indicated as well.
\label{fig:1}}
\end{center}
\end{figure}

We have used the heliocentric Keplerian orbital elements and 1$\sigma$ uncertainties of 2015~WZ$_{12}$ to perform a preliminary exploration of 
its short-term dynamical evolution (for technical details see \citealt{2012MNRAS.427..728D,2013MNRAS.432..886D,2014MNRAS.439.2970D}). Our limited 
analysis strongly suggests that its evolution becomes difficult to reconstruct or predict beyond 100 yr. Figure~\ref{fig:1} shows the evolution 
backward and forward in time of several orbital elements and other relevant parameters of 2015~WZ$_{12}$ using initial conditions compatible with 
the nominal orbit presented above. The top panel shows that 2015~WZ$_{12}$ experiences close encounters with the Earth at relatively short-range; 
these are often low-velocity flybys as they take place at aphelion. The second to top panel shows the behavior of the so-called Kozai-Lidov 
parameter that measures the evolution of the component of the orbital angular momentum perpendicular to the ecliptic; the value remains fairly 
constant. The third to top panel shows the variation of its relative mean longitude; when its value oscillates, the object is engaged in a 1:1 
mean motion resonance. Asteroid 2015~WZ$_{12}$ might have been until recently a transient Trojan of Venus (it lost this status after suffering a 
close encounter with Mercury). Other control orbits show a somewhat different evolution, but in all cases we observe frequent switching between 
the various co-orbital states and their hybrids (see \citealt{2012MNRAS.427..728D,2013MNRAS.432..886D,2014MNRAS.439.2970D}). The bottom panel 
shows the evolution of the nodal distances of 2015~WZ$_{12}$; flybys with the Earth--Moon system take place at the descending node, while Mercury 
is approached at the ascending node.

An object as small as 2015~WZ$_{12}$ must be a fragment of a larger body, it may have its provenance in the main asteroid belt, but it might have
been produced {\it in situ}, i.e. in the region between the orbits of Mercury and the Earth--Moon system, via super-catastrophic break-ups 
\citep{2016Natur.530..303G}. Follow-up observations of this target in the near future will be difficult, though; it will reach its next favorable
perigee in 2018 November, at a solar elongation of about 85\degr with an apparent magnitude in excess of 26. 

\acknowledgments

We thank S.~J. Aarseth for providing the code used in this research, A.~I. G\'omez de Castro, I. Lizasoain and L. Hern\'andez Y\'a\~nez of the 
Universidad Complutense de Madrid (UCM) for providing access to computing facilities. This work was partially supported by the Spanish 
`Ministerio de Econom\'{\i}a y Competitividad' (MINECO) under grant ESP2014-54243-R. Part of the calculations and the data analysis were 
completed on the EOLO cluster of the UCM. EOLO, the HPC of Climate Change of the International Campus of Excellence of Moncloa, is funded by the 
MECD and MICINN. This is a contribution to the CEI Moncloa. In preparation of this paper, we made use of the NASA Astrophysics Data System and 
the MPC data server.

\end{document}